\documentstyle[amsfonts,amssymb,prl,aps,multicol]{revtex}

\newcommand{\ket}[1]{\left | \, #1 \right \rangle}
\newcommand{\bra}[1]{\left \langle #1 \, \right |}

\begin{document}


\onecolumn

\title{Simple Proof of Security of the BB84 Quantum 
Key Distribution Protocol}

\author{Peter W. Shor$^{(1)}$ and John Preskill$^{(2)}$}

\address{$^{(1)}$AT\&T Labs Research, Florham Park, NJ 07932, USA\\ 
$^{(2)}$Lauritsen
Laboratory of High Energy Physics, California Institute of Technology, 
Pasadena, CA 91125, USA }

\date{\today}

\maketitle

\begin{abstract}
We prove that the 1984 protocol of Bennett and Brassard (BB84)
for quantum key distribution is secure.  
We first give a key distribution protocol based on entanglement purification, 
which can be proven secure using methods from Lo and Chau's proof
of security for a similar protocol.  We then show that the security of 
this protocol implies the security of BB84.  The entanglement-purification 
based protocol uses Calderbank-Shor-Steane (CSS) codes, and properties of 
these codes are used to remove the use of quantum computation from the 
Lo-Chau protocol.
\end{abstract}


\begin{multicols}{2}[]


Quantum cryptography differs from conventional cryptography in that 
the data are kept secret by the properties of quantum mechanics, rather 
than the conjectured difficulty of computing certain functions.
The first quantum key distribution protocol, proposed in 1984~\cite{BB}, 
is called BB84 after its inventors (C. H. Bennett and G.~Bras\-sard).
In this protocol, the participants (Alice and Bob)
wish to agree on a secret key about which no
eavesdropper (Eve) can obtain significant information.  
Alice sends each bit of
the secret key in one of a set of conjugate bases which 
Eve does not know, and this key is protected by the impossibility of
measuring the state of a quantum system simultaneously in 
two conjugate bases.  
The original papers proposing quantum key distribution~\cite{BB} proved 
it secure against certain attacks,
including those feasible using
current experimental techniques.  However, for many years, 
it was not rigorously proven secure
against an adversary able to perform any physical
operation permitted by quantum mechanics.

Recently, three proofs of the security of quantum key distribution 
protocols have been discovered; however, none is entirely 
satisfactory.  One proof \cite{LoChau},
although easy to understand, has the drawback that the
protocol requires a quantum computer.
The other two \cite{Mayers,BBBMR} prove the
security of a protocol based 
on BB84, and so 
are applicable to near-practical settings.  However, 
both proofs are quite complicated.
We give a simpler proof by relating the security of BB84 to entanglement
purification protocols \cite{epps} and quantum error correcting 
codes~\cite{CSS}.  
This new proof also may illuminate some 
properties of previous proofs \cite{Mayers,BBBMR},
and thus give insight into them.  For example, it elucidates
why the rates obtainable from these proofs are related to rates
for CSS codes.  The proof was in fact inspired by the
observation that CSS codes are hidden in the inner workings of the proof
given in \cite{Mayers}.


We first review CSS codes and associated entanglement purification
protocols.  
Quantum error-correcting codes are subspaces of the Hilbert 
space~${\Bbb C}^{2^n}$ which are protected from errors in a small 
number of these qubits, so that any such error can be measured and 
subsequently corrected without disturbing the encoded state.  
A quantum CSS code $Q$ on $n$ qubits comes from two binary codes on $n$ 
bits, $C_1$ and $C_2$, one contained in the other: 
\[
\{0\} \subset {C_2} \subset {C_1} \subset {\bf F}_2^n,
\]
where ${\bf F}_2^n$ is the binary vector space on $n$ bits \cite{CSS}.

A set of basis states (which we call {\em codewords}) for the CSS code 
subspace can be obtained from vectors $v \in C_1$ as follows:
\begin{equation}
{v}\, \longrightarrow\, \frac{1}{|C_2|^{1/2}} \sum_{w \in C_2} \ket{v+w}.
\label{codewords}
\end{equation}
If $v_1 - v_2 \in C_2$, then the codewords corresponding
to $v_1$ and $v_2$ are the same.  Hence these codewords 
correspond to cosets of $C_2$ in $C_1$, and this code protects
a Hilbert space of dimension $2^{\dim C_1 - \dim C_2}$.

The above quantum code is equivalent to the dual code $Q^*$ obtained from 
the two binary codes
\[
\{0\} \subset C_1^\perp \subset C_2^\perp \subset {\bf F}_2^n.
\]
This equivalence can be demonstrated by applying the Hadamard transform
\[
H = \frac{1}{\sqrt{2}} \left(\begin{array}{rr}1 & 1 \\ 
1 & -1 \end{array} \right)
\]
to each encoding qubit.  This transformation interchanges the bases
$\ket{0}$, $\ket{1}$ and $\ket{+}$, $\ket{-}$, where 
$\ket{+} = \frac{1}{\sqrt{2}}(\ket{0} + \ket{1})$ and
$\ket{-} = \frac{1}{\sqrt{2}}(\ket{0} - \ket{1})$.
It also interchanges the two subspaces corresponding to the codes $Q$
and $Q^*$, although the codewords 
(given by Eq.~\ref{codewords}) of $Q$ and
$Q^*$ are not likewise interchanged.

We now make a brief technical detour to define some terms.  The 
three Pauli matrices are:
\[
\sigma_x = \left(\begin{array}{rr}0 & 1 \\ 1 & 0 \end{array} \right),
\quad
\sigma_y = \left(\begin{array}{rr}0 & -i \\ i & 0 \end{array} \right),
\quad
\sigma_z = \left(\begin{array}{rr}1 & 0 \\ 0 & -1 \end{array} \right).
\]
The matrix $\sigma_x$ applies a bit flip error to a qubit, while
$\sigma_z$ applies a phase flip error.
We denote the Pauli matrix $\sigma_a$ acting 
on the $k$'th bit of the CSS code
by $\sigma_{a(k)}$ for $a \in \{x,y,z\}$.
For a binary vector~$s$, we let 
\[
\sigma_a^{[s]} = 
\sigma_{a(1)}^{s_1}
\otimes
\sigma_{a(2)}^{s_2}
\otimes
\sigma_{a(3)}^{s_3} 
\otimes\ldots\otimes
\sigma_{a(n)}^{s_n}
\]
where $\sigma_a^0$ is the identity matrix and $s_i$ is the $i$'th 
bit of~$s$.
The matrices $\sigma_x^{[s]}$ ($\sigma_z^{[s]}$) have 
all eigenvalues $\pm 1$.

In a classical error correcting code, correction proceeds
by measuring the syndrome, which is done as follows.  
A {\em parity check} matrix $H$ of a code $C$ is a basis of the dual  
vector space $C^\perp$.  Suppose that we transmit a codeword $v$,
which acquires errors to become $w = v + \epsilon$.  
The $k$'th row $r_k$ of the matrix $H$ determines the $k$'th bit of the 
syndrome for $w$, namely $r_k \cdot w \hbox{\ (mod 2)}$.
The full syndrome is thus $Hw$.  If the syndrome is~$0$, then $w \in C$.  
Otherwise, the most likely value of the error $\epsilon$
can be calculated from the syndrome \cite{Difficult}. 
In our quantum CSS code, we need to correct both
bit and phase errors.  
Let $H_1$ be a parity check matrix for the code $C_1$, and $H_2$ one for 
the code $C_2^\perp$.  
To calculate the syndrome for bit flips,
we measure the eigenvalue of $\sigma_z^{[r]}$ for each row~$r \in H_1$
($-1$'s and $1$'s of the eigenvalue correspond 
to 1's and 0's of the syndrome).
To calculate the syndrome for phase flips, we measure the eigenvalue 
of $\sigma_x^{[r]}$ for each row~$r \in H_2$.
This lets us correct both bit and phase flips, and if we can correct
up to $t$ of each of these types of errors, we can also correct
arbitrary errors on up to $t$ qubits~\cite{CSS}.

The useful property of CSS codes for demonstrating the security of BB84
is that the error correction for the phases is decoupled from 
that for the bit values, as shown above.  General quantum stabilizer
codes can similarly be turned into key distribution protocols, but these
appear to require a quantum computer to implement.

If one requires that a CSS code correct all errors on at most
$t = \delta n$ qubits, the best codes that we know exist satisfy the 
quantum Gilbert-Varshamov bound.  As the block length $n$ 
goes to infinity, these codes asymptotically protect against 
$\delta n$ bit errors and $\delta n$ phase errors, and encode
$[1 - 2 H(2\delta)]n$ qubits, where $H$ is the binary Shannon entropy  
$H(p) = -p \log_2(p) -(1-p) \log_2(1-p)$. 
In practice, 
it is better to only require that random errors are corrected with high
probability.  In this case, codes exist that 
correct $\delta n$ random phase errors and $\delta n$
random bit errors, and which encode $[1-2 H(\delta ) ] n$ qubits.


We also need a description of the Bell basis.
These are the four maximally entangled states
\[
\Psi^\pm = \frac{1}{\sqrt{2}}( \ket{01} \pm \ket{10}),
\quad
\Phi^\pm = \frac{1}{\sqrt{2}}( \ket{00} \pm \ket{11}),
\]
which form an orthogonal basis for the quantum state space of two qubits.

Finally, we introduce a class of quantum error correcting codes equivalent
to $Q$, and parameterized by two $n$-bit binary vectors
$x$ and $z$.  Suppose that $Q$ is determined as above by $C_1$ and $C_2$.
Then $Q_{x,z}$ has basis vectors indexed by cosets of $C_2$ in $C_1$,
and for $v \in C_1$, the corresponding codeword is
\begin{equation}
{v} \,\longrightarrow\, \frac{1}{|C_2|^{1/2}}
\sum_{w \in C_2} (-1)^{z\cdot w} \ket{x+v+w}.
\label{codewords2}
\end{equation}
  

Quantum error correcting codes and entanglement purification protocols
are closely connected \cite{epps}; we now describe the entanglement 
purification protocol corresponding to the CSS code $Q$.
For now, we assume that the codes
$C_1$ and $C_2^\perp$ correct up to $t$ errors and 
that $Q$ encodes $m$ qubits in $n$ qubits.
Suppose Alice and Bob share $n$ pairs
of qubits in a state close to $(\Phi^+)^{\otimes n}$.
For the entanglement purification 
protocol, Alice and Bob separately measure the eigenvalues of
$\sigma_z^{[r]}$ for each row $r \in H_1$ and 
$\sigma_x^{[r']}$ for each row $r' \in H_2$.   
Note that for these measurements to be performable simultaneously, they
must all commute;  $\sigma_z^{[r]}$ and $\sigma_x^{[r']}$
commute because the vector spaces
$C_1^\perp$ and $C_2$ are orthogonal.

If Alice and Bob start with $n$ perfect EPR pairs, measuring
$\sigma_z^{[r]}$ for $r \in H_1$ and $\sigma_x^{[r']}$ for $r' \in H_2$
projects each of their states onto the code subspace $Q_{x,z}$, 
where $x$ and $z$ are any binary vectors with
$H_1 x$ and $H_2 z$ equal to the measured bit and phase syndromes,
respectively.
After projection, the state is $(\Phi^{+})^{\otimes m}$ encoded by $Q_{x,z}$.

Now, suppose that Alice and Bob start with a state close 
to $(\Phi^+)^{\otimes n}$.
To be specific, suppose that all their EPR pairs are in the Bell basis,
with $t$ or fewer bit flips ($\Psi^+$ or $\Psi^-$ pairs)
and $t$ or fewer phase flips ($\Phi^-$ or $\Psi^-$ pairs).  
If Alice and Bob compare their measurements of $\sigma_z^{[r]}$ 
($\sigma_x^{[r]}$), the rows $r$ for which these measurements disagree
give the bits which are $1$ in the bit (phase) syndromes.  
 From these syndromes, Alice and Bob can compute the locations of the bit 
and the phase flips, can correct these errors, and can
then decode $Q_{x,z}$ to obtain $m$ perfect EPR pairs.

We will show that the following is a secure quantum key 
distribution protocol.

\vspace{\baselineskip}

\centerline{{\bf Protocol 1: Modified Lo-Chau}}
\begin{itemize}
\setlength{\itemsep}{-\parskip}
\item [\bf 1:] Alice creates $2n$ EPR pairs in 
        the state $(\Phi^+)^{\otimes n}$.
\item [\bf 2:] Alice selects a random $2n$ bit string $b$, and performs
        a Hadamard transform on the second half of each EPR pair for
        which $b$ is 1.
\item [\bf 3:] Alice sends the second half of each EPR pair to Bob.
\item [\bf 4:] Bob receives the qubits and publicly announces this fact.
\item [\bf 5:] Alice selects $n$ of the $2n$ encoded EPR pairs to serve as
	check bits to test for Eve's interference.  
\item [\bf 6:] Alice announces the bit string $b$, and which $n$ EPR pairs
        are to be check bits.
\item [\bf 7:] Bob performs Hadamards on the qubits where $b$ is~$1$.
\item [\bf 8:] Alice and Bob each measure their halves of the $n$ check EPR
	pairs in the $\ket{0}$, $\ket{1}$ basis and share the results.  
        If too many of these measurements disagree, they abort the
        protocol.  
\item [\bf 9:] Alice and Bob make the measurements on their code qubits
        of $\sigma_z^{[r]}$ for each row $r \in H_1$ and $\sigma_x^{[r]}$ for
        each row $r \in H_2$.  Alice and Bob share the results, compute
        the syndromes for bit and phase flips, and then transform their 
        state so as to obtain $m$ nearly perfect EPR pairs.  
\item [\bf 10:] Alice and Bob measure the EPR pairs in the 
        $\ket{0}$, $\ket{1}$ basis to obtain a shared secret key.
\end{itemize}


We now show that this protocol works.  Namely, we show that the probability 
is exponentially small that Alice and Bob agree on a key about which Eve can 
obtain more than an exponentially small amount of information.
We need a result of Lo and Chau
\cite{LoChau} that if Alice and Bob share a state having fidelity
$1-2^{-s}$ with $(\Phi^+)^{\otimes m}$, then Eve's mutual information
with the key is at most $2^{-c} + 2^{O(-2s)}$ where 
$c = s-\log_2 ( 2m +s + 1/\log_e 2 )$.

For the proof, we use an argument
based on one from Lo and Chau \cite{LoChau}.  Let us
calculate the probability that the test on the check bits succeeds while the
entanglement purification on the code bits fails.  We do this by 
considering the measurement that projects each of the EPR pairs onto the 
Bell basis.

We first consider the check bits.  Note that for the EPR pairs where $b = 1$,
Alice and Bob are effectively measuring them in the $\ket{+}$, $\ket{-}$
basis rather than the $\ket{0}$, $\ket{1}$ basis.  Now, observe that
\begin{eqnarray*}
\ket{\Psi^+} \bra{\Psi^+} + \ket{\Psi^-} \bra{\Psi^-} 
&=& \ket{01\,} \bra{\,01} + \ket{10\,} \bra{\,10}, \\
\ket{\Phi^-} \bra{\Phi^-} + \ket{\Psi^-} \bra{\Psi^-} 
&=& \ket{+-} \bra{+-} + \ket{-+} \bra{-+}.
\end{eqnarray*}
These relations show that the rates of bit flip errors and
of phase flip errors that Alice and Bob estimate from their measurements on
check bits are the same as they would have estimated using the Bell basis
measurement.

We next consider the measurements on the code bits.   We want to show that the
purification protocol applied to $n$ pairs produces a state that is close to
the encoded $(\Phi^+)^{\otimes m}$. The purification protocol succeeds
perfectly acting on the space spanned by Bell pairs that differ from
$(\Phi^+)^{\otimes n}$ by $t$ or fewer bit flip errors and by $t$ or fewer
phase flips errors. Let $\Pi$ denote the projection onto this space. Then if
the protocol is applied to an initial density operator $\rho$ of the $n$ pairs,
it can be shown that the final density operator $\rho'$ approximates
$(\Phi^+)^{\otimes m}$ with fidelity 
\begin{equation}
F \equiv \langle (\Phi^+)^{\otimes m}|~\rho' ~|(\Phi^+)^{\otimes m}\rangle \ge
{\rm tr} \left (\Pi \rho \right) ~.
\end{equation}
Hence the fidelity is at least as large as the probability that $t$ or fewer
bit flip errors and $t$ or fewer phase flip errors would have been found, if
the Bell measurement had been performed on all $n$ pairs.

Now, when Eve has access to the qubits, she does not yet know 
which qubits are check qubits and which are code qubits, so she cannot
treat them differently.
The check qubits that Alice and Bob measure thus behave like a classical 
random sample of the qubits.  We are then able to use the measured 
error rates in a classical probability estimate; we find that 
probability of obtaining more than $\delta n$ 
bit (phase) errors on the code bits and fewer than $(\delta - \epsilon)n$ 
errors on the check bits is asymptotically less than
$\exp[-\frac{1}{4}\epsilon^2 n / (\delta - \delta^2) ]$.
We conclude that if Alice and Bob have greater than an exponentially
small probability of passing the test, then the fidelity of Alice and Bob's 
state with $(\Phi^+)^{\otimes m}$ is exponentially close to 1.

We now show how to turn this Lo-Chau type protocol into a quantum
error-correcting code protocol.  Observe first that it does not matter 
whether Alice
measures her check bits before or after she transmits half of each EPR
pair to Bob, and similarly that it does not matter whether she measures
the syndrome before or after this transmission.  If she measures the check
bits first, this is the same as choosing a random one of $\ket{0}$,
$\ket{1}$. 
If she also
measures the syndrome first, this is equivalent to transmitting $m$ halves 
of EPR pairs encoded by the CSS code $Q_{x,z}$ for two random 
vectors $x$, $z \in {\bf F}_2^n$.  The vector $x$ is determined by the 
syndrome measurements $\sigma_z^{[r]}$ for rows $r \in H_1$, and similarly 
for $z$.
Alice can also measure her half of the 
encoded EPR pairs before or after transmission.  If she measures them
first, this is the same as 
choosing a random key $k$ and encoding $k$ using ${Q}_{x,z}$.
We thus obtain the following equivalent protocol.

\vspace{\baselineskip}

\centerline{{\bf Protocol 2: CSS Codes}}
\begin{itemize}
\setlength{\itemsep}{-\parskip}
\item [\bf 1:] Alice creates $n$ random check bits, 
        a random $m$-bit key $k$, and a random $2n$-bit string $b$.

\item [\bf 2:] Alice chooses $n$-bit strings $x$ and $z$ at random. 

\item [\bf 3:] Alice encodes her key $\ket{k}$ using the CSS code
        $Q_{x,z}$

\item [\bf 4:] Alice chooses $n$ positions (out of $2n$) and puts 
        the check bits in these positions and the code bits in the 
        remaining positions.

\item [\bf 5:] Alice applies a Hadamard transform to those qubits
        in the positions having 1 in $b$.  

\item [\bf 6:] Alice sends the resulting state to Bob.  Bob acknowledges
        receipt of the qubits.

\item [\bf 7:] Alice announces $b$, the positions of the check bits, 
        the values of the check bits, and the $x$ and $z$ determining
        the code $Q_{x,z}$.

\item [\bf 8:] Bob performs Hadamards on the qubits where $b$ is $1$. 

\item [\bf 9:] Bob checks whether too many of the check bits have
        been corrupted, and aborts the protocol if so.

\item [\bf 10:] Bob decodes the key bits and uses them for the key.

\end{itemize}


Intuitively, the security of the protocol depends on the fact
that for a sufficiently low error rate, a CSS code transmits the 
information encoded by it with very high fidelity, so that by the 
no-cloning principle very little information can leak to Eve.

We now give the final argument that turns the above protocol into BB84.  
First note that, since all Bob cares about are the bit values of the 
encoded key, and the string $z$ is only used to correct the phase of 
the encoded qubits, Bob does not need $z$.  This is why we use CSS codes: 
they decouple the phase correction from the bit correction.   
Let $k' \in C_1$ be a binary vector that is mapped by 
Eq.~(\ref{codewords2}) to the encoded key. 
Since Bob never uses $z$, we can assume that Alice does not send it.  
Averaging over $z$, we see that Alice effectively
sends the mixed state
\begin{eqnarray}
\nonumber
\frac{1}{2^n|C_2|} 
\sum_z \Big[ \sum_{w_1, w_2 \in C_2} (-1)^{(w_1+w_2) \cdot z} \hspace*{1in}\\
\hspace*{1in}\times\ket{k'+w_1+x}\bra{k'+w_2+x}\Big] \nonumber \\
= \frac{1}{|C_2|} \sum_{w \in C_2} \ket{k'+w+x}\bra{k'+w+x},\hspace*{.2in}
\end{eqnarray}
which is equivalently the mixture of states $\ket{k'+x+w}$ with $w$ chosen 
randomly in $C_2$.  Let us now look at the protocol as a whole.  
The error correction information 
Alice gives Bob is $x$, and Alice sends $\ket{k'+x+w}$ over the quantum
channel.  Over many
iterations of the algorithm, these are random variables chosen uniformly
in ${\bf F}_2^n$ with the constraint that their difference $k' + w$ is
in~$C_1$.  After Bob receives $k'+w+x+\epsilon$, he subtracts $x$, and 
corrects the result to a codeword in $C_1$, which is almost 
certain to be $k'+w$.  The key is the coset of $k'+w$ over~$C_2$.  

In the BB84 protocol given below, Alice sends $\ket{v}$ to Bob,
with error correction information $u + v$.   These are again two random
variables uniform in ${\bf F}_2^n$, with 
the constraint that $u\in C_1$.  Bob obtains
$v+\epsilon$, subtracts $u+v$, and corrects the result to a
codeword in $C_1$, which with high probability is $u$.
The key is then the coset $u + C_2$.
Thus, the two protocols are completely equivalent.

\vspace{\baselineskip}

\centerline{{\bf Protocol 3: BB84}}
\begin{itemize}
\setlength{\itemsep}{-\parskip}

\item [\bf 1:] Alice creates $(4+\delta)n$ random bits.

\item [\bf 2:] Alice chooses a random $(4+\delta)n$-bit string $b$.  For 
	each bit, she creates a state in the $\ket{0}$, $\ket{1}$
	basis (if the corresponding bit of $b$ is $0$) or the $\ket{+}$, 
        $\ket{-}$ basis (if the bit of $b$ is $1$).

\item [\bf 3:] Alice sends the resulting qubits to Bob.

\item [\bf 4:] Bob receives the $(4+\delta)n$ qubits, measuring each
	in the $\ket{0}$,$\ket{1}$ or the $\ket{+}$,$\ket{-}$ 
        basis at random.

\item [\bf 5:] Alice announces $b$.

\item [\bf 6:] Bob discards any results where he measured a different
	basis than Alice prepared.  With high probability, there are at
	least $2n$ bits left (if not, abort the protocol).  Alice 
        decides randomly on a set of $2n$ bits to use for the protocol, and
        chooses at random $n$ of these to be check bits. 

\item [\bf 7:] Alice and Bob announce the values of their check bits.
	If too few of these values agree, they abort the protocol.

\item [\bf 8:] Alice announces $u+v$, where $v$ is the string
        consisting of the remaining non-check bits, and $u$ is a random 
        codeword in $C_1$.

\item [\bf 9:] Bob subtracts $u+v$ from his code qubits, $v+\epsilon$, and
        corrects the result, $u+\epsilon$, to a codeword in $C_1$.  

\item [\bf 10:] Alice and Bob use the coset of $u + C_2$ as the key.

\end{itemize}



There are a few loose ends that need to be tied up.  The protocol
given above uses binary codes $C_1$ and $C_2^\perp$ with large
minimum distance, and thus can obtain rates given by the quantum
Gilbert-Varshamov bound for CSS codes \cite{CSS}.  To reach the better
Shannon bound for CSS codes, we need to use codes for which a random 
small set of phase errors and bit errors can almost always be corrected.  
To prove that the protocol works in this case, we need to ensure that the 
errors are indeed random.  We do this by adding a step where Alice 
scrambles the qubits using a random permutation $\pi$ before sending 
them to Bob, and a step after Bob acknowledges receiving the qubits 
where Alice sends $\pi$ to Bob and he unscrambles the qubits.   
This can work as long as the measured bit and phase 
error rates are less than 11\%, the point at which the Shannon rate 
$1-2H(\delta)$ hits $0$.

For a practical key distribution protocol we 
need the classical code $C_1$ to be efficiently decodeable.
As is shown in \cite{Mayers}, we can let $C_2$ be a random subcode of 
an efficiently decodeable code $C_1$, and with high probability obtain 
a good code $C_2^\perp$.  While known efficiently decodeable codes do 
not meet the Shannon bound, they come fairly close. 

A weakness in both the proof given in this paper and the proofs
in \cite{Mayers,BBBMR} is that they do not apply to imperfect sources;
the sources must be perfect single-photon sources.  A proof avoiding
this difficulty was recently discovered by Michael Ben-Or \cite{BenOr};
it shows that any source 
sufficiently close to a single-photon source is still secure.   
However, most experimental quantum key distribution systems 
use weak coherent sources, and no currently known proof covers this case.

The authors thank Michael Ben-Or, Eli Biham, Hoi-Kwong Lo, 
Dominic Mayers and Tal Mor for explanations of and informative discussions 
about their security proofs.  We also thank Ike Chuang, 
Dan Gottesman, Alexei Kitaev and Mike Nielsen for their 
discussions and suggestions, which greatly improved this paper.  Part of 
this research was done while PWS was visiting Caltech.  This work
has been supported in part by the Department of Energy under Grant
No.\ DE-FG03-92-ER40701, and by DARPA through Caltech's Quantum Information 
and Computation (QUIC) project administered by the Army Research Office.

\bibliographystyle{prsty}

\begin{thebibliography}{1}
\bibitem{Difficult}
This calculation may be quite difficult, but for now we ignore this 
practical complication.
\end{thebibliography}

\end{multicols}
\end{document}